\documentclass[12pt,preprint]{aastex}
\newcommand{\gtorder}{\mathrel{\raise.3ex\hbox{$>$}\mkern-14mu
             \lower0.6ex\hbox{$\sim$}}}
\newcommand{\ltorder}{\mathrel{\raise.3ex\hbox{$<$}\mkern-14mu
             \lower0.6ex\hbox{$\sim$}}}

\begin{document}
\begin{abstract}

Mergers between stellar-mass black holes will be key sources of
gravitational radiation for ground-based detectors.  However, the rates
of these events are highly uncertain, given that such systems are
invisible.  One formation scenario involves mergers in field binaries,
where our lack of complete understanding of common envelopes and the
distribution of supernova kicks has led to rate estimates that range
over a factor of several hundred.  A different, and highly promising,
channel involves multiple encounters of binaries in globular clusters or
young star clusters.  However, we currently lack solid evidence for
black holes in almost all such clusters, and their low escape speeds
raise the possibility that most are ejected because of supernova
recoil.  Here we propose that a robust environment for mergers
could be the nuclear star clusters found in the centers of small
galaxies.  These clusters have millions of stars, black hole relaxation
times well under a Hubble time, and escape speeds that are several times
those of globulars, hence they retain most of their black holes. 
We present simulations of the three-body dynamics of black
holes in this environment and estimate that, if most nuclear star clusters
do not have supermassive black holes that interfere with the mergers, 
at least several tens of events per year will be detectable with Advanced 
LIGO.

\end{abstract}


\title{Mergers of Stellar-Mass Black Holes in Nuclear Star Clusters}
\author{M. Coleman Miller and Vanessa M. Lauburg}
\affil{University of Maryland, Department of Astronomy,
College Park, Maryland 20742-2421}
\maketitle

\section{Introduction}

Ground-based gravitational wave detectors have now achieved their
initial sensitivity goals (e.g., \citealt{Abbott07}).  
In the next few years, these sensitivities
are expected to improve by a factor of $\sim 10$, 
which will increase the searchable volume by a factor of $\sim 10^3$ 
and will lead to many detections per year.

One of the most intriguing possible sources for such detectors is
the coalescence of a double stellar-mass black hole binary.  Such
binaries are inherently invisible, meaning that we have no direct
observational handle on how common they are or their masses, spin
magnitudes, or orientations.  Comparison of the observed waveforms
(or of waveforms from merging supermassive black holes) with
predictions based on post-Newtonian analysis and numerical
relativity will be the most direct possible test of the
predictions of strong-gravity general relativity.

The electromagnetic non-detection of these sources makes
rate estimates highly challenging, because our only observational
handles on BH-BH binaries come from possible progenitors.  For example, a
common scenario involves the effectively isolated evolution of a
field binary containing two massive stars into a binary with two black
holes that will eventually merge (e.g., \citealt{Lipunov1997,BelBul1999}).  
There are profound uncertainties
involved in calculations of these rates due to e.g., the lack of 
knowledge of the details of the common envelope phase in these
systems and the absence of guides to the distribution of supernova kicks
delivered to black holes.  As a recent indication of the range of estimated
rates, note that the Advanced LIGO detection rate of BH-BH coalescences 
is estimated to be anywhere between $\sim 1-500$~yr$^{-1}$
by \cite{Belczynski2007}, depending on how common envelopes are modeled.  

Another promising location for BH-BH mergers is globular clusters or
super star clusters,  where stellar number densities are high enough
to cause multiple encounters and hardening of binaries. Even though
binaries are kicked out before they merge 
\citep{KHM93,SH93,SP93,SP95,PZMcM2000,OLeary2006}, these clusters can still
serve as breeding grounds for gravitational wave sources.  Indeed,
\cite{OLeary2007} estimate a rate of 0.5~yr$^{-1}$  for initial LIGO
and 500~yr$^{-1}$ for Advanced LIGO via this channel.  There is,
however, little direct evidence for black holes in most globulars
(albeit they could be difficult to see).  In addition, at least one
black hole in a low-mass X-ray binary apparently received a $\gtorder
100$~km~s$^{-1}$ kick from its supernova (GRO~J1655--40; see
\citealt{Mir02}).  This is double the escape speed from the centers
of even fairly rich globulars \citep{Webbink1985},  leading to
uncertainties about their initial black hole population and  current
merger rates.

Here we propose that mergers occur frequently in the nuclear star
clusters that may be in the centers of many low-mass galaxies
(\citealt{Boecker2002,Ferrarese2006,WH06}; note that some of these
are based on small deviations from smooth surface brightness profiles and
are thus still under discussion).  It has recently been recognized that
in these galaxies, which may not have supermassive black holes
(for a status report on ongoing searches for low-mass central 
black holes, see \citealt{GH07}), the
nuclear clusters have masses that are correlated with the
surrounding velocity dispersion $\sigma$ as $M\approx 10^7~M_\odot 
(\sigma/54~{\rm km~s}^{-1})^{4.3}$ \citep{Ferrarese2006}.   When the
velocity dispersion is in the range of $\sim 30-60$~km~s$^{-1}$, the
half-mass relaxation time is  small enough that black holes (which
have $\sim 20\times$ the average stellar mass) can sink to the
center in much less than a Hubble time.  In addition, although
systems with equal-mass objects require roughly 15 half-mass
relaxation times to undergo core collapse \citep{BinTre1987},
studies show that systems with a wide range of stellar masses
experience core collapse within $\sim 0.2\times$ the half-mass
relaxation time \citep{PZMcM2002,Gurkan2004}.   Therefore, clusters
with masses less than $\sim {\rm few}\times 10^7~M_\odot$ will have
collapsed  by now and hence increased the escape speed from the
center, allowing retention of most of their black holes.

As we show in this paper, nuclear star clusters are therefore excellent
candidates for stellar-mass black hole binary mergers because they keep
their black holes while also evolving rapidly enough that the holes can
sink to a region of high density.  If tens of percent of the black holes
in eligible galaxies undergo such mergers, the resulting rate for Advanced
LIGO is at least several tens per year.  In \S~2
we quantify these statements and results more precisely and discuss our
numerical three-body  method.  We give our conclusions in \S~3.

\section{Method and Results}

\subsection{Characteristic times and initial setup}

Our approach is similar to that of \cite{OLeary2006}, who
focus on globular clusters with velocity dispersions 
$\sigma\leq 20$~km~s$^{-1}$.  Here, however, we concentrate on
the more massive and tightly bound nuclear star clusters.
Our departure point is the relation found by \cite{Ferrarese2006}
between the masses and velocity dispersions of such
clusters:
\begin{equation}
M_{\rm nuc}=10^{6.91\pm 0.11}
\left(\sigma/54~{\rm km~s}^{-1}\right)^{4.27\pm 0.61}~M_\odot\; .
\end{equation}
Assuming that there is no massive central black hole for these
low velocity dispersions, the half-mass relaxation time for the system is
(see \citealt{BinTre1987})
$t_{\rm rlx}\approx {N/2\over{8\ln N}}t_{\rm cross}$
where $N\approx M_{\rm nuc}/0.5~M_\odot$ is the number of stars
in the system (assuming an average mass of $0.5~M_\odot$) 
and $t_{\rm cross}\approx R/\sigma$ is the crossing
time.  Here $R=GM_{\rm nuc}/\sigma^2$ is the radius of the cluster.
Putting this together gives
\begin{equation}
t_{\rm rlx}\approx 1.3\times 10^{10}~{\rm yr}
\left(\sigma/54~{\rm km~s}^{-1}\right)^{5.54\pm 0.61}\; .
\end{equation}
The relaxation time scales inversely with the mass of an individual
star \citep{BinTre1987}, so a $10~M_\odot$ black hole will settle 
in roughly 1/20 of
this time.  Also note that large N-body simulations with broad
mass functions evolve to core collapse within roughly 0.2 half-mass
relaxation times \citep{PZMcM2002,Gurkan2004}, hence in 
the current universe clusters with
velocity dispersions $\sigma<60$~km~s$^{-1}$ will have had their
central potentials deepened significantly.  

The amount of deepening of the potential, and thus the escape speed
from the center of the cluster, depends on uncertain details such
as the initial radial dependence of the density and the binary
fraction.  Given that the timescale for
segregation of the black holes in the center is much less than a
Hubble time, we will assume that the escape speed is roughly
$5\times$ the velocity dispersion, as is the case for relatively
rich globular clusters \citep{Webbink1985}.  This may well be
somewhat conservative, because the higher velocity dispersion here
than in globulars suggests that a larger fraction of binaries will 
be destroyed in nuclear star clusters.  This could lead to less
efficient central energy production and hence deeper
core collapse than is typical in globulars.

With this setup, our task is to follow the interactions of black
holes in the central regions of nuclear star clusters, where we will
scale by stellar number densities of $n\sim 10^6$~pc$^{-3}$ because
of density enhancements caused by relaxation and mass segregation. 
Some black holes will begin their lives in binaries, but to be
conservative we will assume that they start as single objects and
have to exchange into binaries that contain main sequence stars or
other objects.  All binaries in the cluster will be hard, i.e., will
have internal energies greater than the average kinetic energy of a
field star, because otherwise they will be softened and ionized
quickly (e.g., \citealt{BinTre1987}).   If, for example, we consider
binaries of two $1~M_\odot$ stars in a system with
$\sigma=50$~km~s$^{-1}$,  this means that the semimajor axis has to
be less than $a_{\rm max}\sim 1$~AU.  Studies of main sequence
binaries in globular clusters, which have $\sigma\sim
10$~km~s$^{-1}$,  suggest that after billions of years roughly
5--20\% of them survive, with the rest falling victim to ionization
or collisions \citep{Iva05}.  The binary fraction will be lower in
nuclear star clusters due to their enhanced velocity dispersion, but
since when binaries are born they appear to have a constant
distribution across the log of the semimajor axis from $\sim
10^{-2}-10^3$~AU (e.g., \citealt{Abt83,Duquennoy91}) the reduction is
not necessarily by a large factor.  We will conservatively  scale by
a binary fraction $f_{\rm bin}=0.01$.

If a black hole with mass $M_{\rm BH}$ gets within a couple of
semimajor axes of a main sequence binary, the binary will tidally separate and
the BH will acquire a companion.  The timescale on which this
happens is $t_{\rm bin}=(n\Sigma\sigma)^{-1}$, where
$\Sigma=\pi r_p^2\left(1+2GM_{\rm tot}/(\sigma^2 r_p)\right)$ is
the interaction cross section for pericenter distances
$\leq r_p$ when gravitational focusing is included.  Here 
$M_{\rm tot}$ is the mass of the black hole plus the mass of the binary.
If we assume that $M_{\rm BH}=10~M_\odot$ and it interacts with a
binary with two $1~M_\odot$ members and an $a=1$~AU semimajor axis,
then the typical timescale on which a three-body interaction
and capture of one of the stars occurs is
\begin{equation}
t_{\rm 3-bod}=(n\Sigma\sigma)^{-1}\approx 1.2\times 10^9~{\rm yr}
(n/10^6~{\rm pc}^{-3})^{-1}(f_{\rm bin}/0.01)^{-1}
(\sigma/50~{\rm km~s})(a/1~{\rm AU})^{-1}\; .
\end{equation}
This is small enough compared to a Hubble time that we start our
simulations by assuming that each black hole has exchanged into a 
hard binary, and follow its evolution from there.

Another important question is whether, after a three-body interaction,
a black hole binary will shed the kinetic energy of its center of mass
via dynamical friction and sink to the center of the cluster before
another three-body encounter.  If not, the kick speeds will add in a
random walk, thus increasing the ejection fraction.

To compute this we note that the local relaxation time of a binary is
\begin{equation}
t_{\rm rlx}={0.339\over{\ln\Lambda}}{\sigma^3\over{
G^2\langle m\rangle M_{\rm bin}n}}
\end{equation}
\citep{Spit87} where $\sigma$ is the local velocity dispersion, 
$\ln\Lambda\sim 10$ is the Coulomb logarithm,
$\langle m\rangle$ is the average mass of interloping stars,
$n$ is their number density, and $M_{\rm bin}$ is the mass of the
binary.  The timescale for a three-body interaction is
$t_{\rm 3-bod}=(n\Sigma\sigma)^{-1}$ as above.
For a gravitationally focused binary, which is of greatest interest
because only these could in principle produce three-body recoil
sufficient to eject binaries or singles, $r_p<GM_{\rm bin}/\sigma^2$.
Therefore, $\Sigma\approx 2\pi r_p GM_{\rm bin}/\sigma^2$ and
\begin{equation}
t_{\rm 3-bod}\approx {\sigma\over{2\pi nr_p GM_{\rm bin}}}\; .
\end{equation}
If we let $r_p=qGM_{\rm bin}/\sigma^2$, with $q<1$, then
\begin{equation}
t_{\rm 3-bod}\approx {\sigma^3\over{2\pi qG^2M_{\rm bin}^2 n}}
\end{equation}
so that
\begin{equation}
t_{\rm rlx}/t_{\rm 3-bod}\approx {2q\over{\ln\Lambda}}{M_{\rm bin}\over{
\langle m\rangle}}\; .
\end{equation}
In the center of a cluster, where mass segregation is likely to
have flattened the mass distribution, we find that this quantity is
typically less than unity (and it decreases as the binary hardens), 
meaning that after a three-body encounter
a binary has an opportunity to share its excess kinetic energy via
two-body encounters and thus settle back to the center of the cluster.
We therefore treat the encounters separately rather than adding
the kick speeds in a random walk.  

In a given encounter, suppose that a binary of total mass 
$M_{\rm bin}=M_1+M_2$, a reduced mass $\mu=M_1M_2/M_{\rm bin}$, and a
semimajor axis $a_{\rm init}$ interacts with an interloper of mass 
$m_{\rm int}$, and that the kinetic energy of the interloper at infinity
is much less than the binding energy of the binary (i.e., this is a
very hard interaction).  If after the interaction the semimajor axis is
$a_{\rm fin}<a_{\rm init}$, then energy and momentum conservation mean 
that the recoil speed of the binary is given by 
$v_{\rm bin}^2=G\mu{m_{\rm int}\over{M_{\rm bin}+m_{\rm int}}}
\left(1/a_{\rm fin}-1/a_{\rm init}\right)$, and the recoil speed of
the interloper is $v_{\rm int}=(M_{\rm bin}/m_{\rm int})v_{\rm bin}$.
For example, suppose that $M_1=M_2=10~M_\odot$, $M_{\rm int}=1~M_\odot$,
$a_{\rm init}=0.1$~AU, and $a_{\rm fin}=0.09$~AU.  The binary then recoils
at $v_{\rm bin}=15$~km~s$^{-1}$ and stays in the cluster, whereas the 
interloper recoils at $v_{\rm int}=300$~km~s$^{-1}$ and is ejected. 

\subsection{Results}

The central regions of the clusters undergo significant
mass segregation, and thus the mass function will be at least
flattened, and possibly inverted.  This has, for example, been observed 
for globulars \citep{Sos97}.  To include this effect, when we consider the
mass of a black hole, its companion, or the interloping third
object in a binary-single encounter, we go through two steps.
First we select a zero age main sequence (ZAMS) mass between
$0.2~M_\odot$ and $100~M_\odot$ using a simple power law distribution
$dN/dM\propto M^{-\alpha}$.  We allow $\alpha$ to range anywhere
from 2.35 (the unmodified Salpeter distribution) to $-1.0$, where
smaller values indicate the effects of mass segregation.
Second, we evolve the ZAMS mass to a current mass.
Our mapping is that for $M_{\rm ZAMS}<1~M_\odot$, the star is still
on the main sequence and retains its original mass; for
$1~M_\odot<M_{\rm ZAMS}<8~M_\odot$ the star has evolved to a white
dwarf, with mass $M_{\rm WD}=0.6~M_\odot+
0.4~M_\odot(M_{\rm ZAMS}/M_\odot-0.6)^{1/3}$; for
$8~M_\odot<M_{\rm ZAMS}<25~M_\odot$ the star has evolved to a 
neutron star, with mass $M_{\rm NS}=1.5~M_\odot+
0.5~M_\odot(M_{\rm ZAMS}-8~M_\odot)/17~M_\odot$; and for
$M_{\rm ZAMS}>25~M_\odot$ the star has evolved to a black hole with
mass $M_{\rm BH}=3~M_\odot+
17~M_\odot(M_{\rm ZAMS}-25~M_\odot)/75~M_\odot$.  Therefore, we
assume that black hole masses range from $3~M_\odot$ to $20~M_\odot$.

These prescriptions are overly simplified in many ways.  We therefore
explore different mass function slopes, main sequence cutoffs, and
so on, and find that our general picture is robust against specific
assumptions.  Note that, consistent with \cite{OLeary2006},
we find that there is a strong tendency for the merged black
holes to be biased towards high masses.  Therefore, if black holes
with masses $>20~M_\odot$ are common, these will dominate the
merger rates.  This is important for data analysis strategies, 
because the low-frequency cutoff of ground-based gravitational wave
detectors implies that higher-mass black holes will have proportionally
more of their signal in the late inspiral, merger, and ringdown.

The three-body interactions themselves are assumed to be Newtonian
interactions between point masses and are computed using the
hierarchical N-body code HNBody (K. Rauch and D. Hamilton, in
preparation), using the driver IABL developed by Kayhan G\"ultekin
(see \citealt{Gultekin2004,Gultekin2006}  for a detailed
description).  These codes use a number of high-accuracy techniques
to follow the evolution of gravitating point masses.  Between
interactions, we use the Peters equations \citep{Peters1964} to
follow the gradual inspiral and circularization of the binary via
emission of gravitational radiation.  This is negligible except near
the end of any given evolution.

We begin by selecting the mass of the black hole and of its companion
(which does not need to be a black hole) from the evolved mass
function.  We also begin with a semimajor axis that is 1/4 of the
value needed to ensure that the binary is hard.  We do this because
soft binaries are likely to be ionized and thus become single stars
rather than merge.  We also select an eccentricity
from a thermal distribution $P(e)de=2ede$.
We then allow the binary to interact with single field
stars drawn from the evolved mass function, one at a time, until
either (1)~the binary merges due to gravitational radiation, 
(2)~the binary is split apart and thus ionized (this is
exceedingly rare given our initial conditions), or (3)~the binary
is ejected from the cluster.  The entire set of interactions until
merger typically takes millions to tens of millions of years, and
only rarely over a hundred million years, so it finishes in much
less than a Hubble time.  In the course of these interactions
there are typically a number of exchanges, which usually swap in
more massive for less massive members of the binary.  This is the
cause of the bias towards high-mass mergers that was also found
by \cite{OLeary2006}.  As shown in Table~1, for $\alpha<1$ most
black holes acquire a black hole companion in the process of
exchanges, and for $\alpha\leq 0.5$ virtually all do.

The results in Table~1 are focused on different mass function
slopes and escape speeds.  As expected, we find that for
$V_{\rm esc}>150$~km~s$^{-1}$ the
overwhelming majority of black hole binaries merge in the nuclear
star cluster rather than being ejected (see Figure~1).  This is the difference
from lower-$\sigma$ globular clusters, where the mergers happen
outside the cluster.  Note also that in addition to few binaries being
ejected, there are typically only 1--2 single black holes ejected per
merger, suggesting that $>50$\% of holes will merge.  In contrast,
at the 50~km~s$^{-1}$ escape speed typical of globulars, $>20$
single black holes are ejected per merger, suggesting an efficiency
of $<10$\%.  For
well-segregated clusters (with $\alpha\leq 0$), the average
mass of black holes that merge, binary ejection fraction and
number of singles ejected, and number of black holes that merge
with each other instead of other objects are all insensitive to
the particular mass function slope.  For less segregated clusters
with $\alpha>0$, the retention fraction of black holes rises rapidly to 
unity because most of the objects that interact with the holes
are less massive stars.  In such cases there might be a channel by
which the mass of the holes increases via accretion of stars, but
we expect $\alpha>0$ to be rare for nuclear star clusters because
of the shortness of the segregation times of black holes.  Overall,
there appears to be a wide range of realistic parameters in which
fewer than 10\% of binary black holes are ejected before merging.

\section{Discussion and Conclusions}

We have shown that nuclear star clusters with velocity dispersions
around $\sigma\sim 30-60$~km~s$^{-1}$ are promising breeding
grounds for stellar-mass black hole mergers.  At significantly
lower velocity dispersions, as found in globulars, the escape
speed is low enough that the binaries are ejected before they
merge.  Significantly higher velocity dispersions appear
correlated with the appearance of supermassive black holes
\citep{Gebhardt2000,Ferrarese2000}.  In such an environment 
there might also be interesting
rates of black hole mergers, but the increasing velocity
dispersion closer to the central object means that binary
fractions are lower and softening, ionization, or tidal
separation by the supermassive black hole itself are strong
possibilities for stellar-mass binaries (\citealt{Miller2005};
Lauburg \& Miller, in preparation).

To estimate the rate of detections with Advanced LIGO, we note
that velocity dispersions in the $\sigma\sim 30-60$~km~s$^{-1}$
range correspond to roughly a factor of $\sim 10$
in galaxy luminosity \citep{Ferrarese2006}.  Galaxy surveys suggest 
(e.g., \citealt{Blan03}) that for dim galaxies
the luminosity function scales
as roughly $dN/dL=\phi^*(L/L_*)^{-\alpha}$, where 
$\phi^*=1.5\times 10^{-2}h^3$~Mpc$^{-3}\approx 5\times 10^{-3}$
Mpc$^{-3}$ for $h=0.71$, and $\alpha\approx -1$.  This implies
that there are nearly equal
numbers of galaxies in equal logarithmic bins of luminosity.
A factor of 10 in luminosity is roughly $e^2$, so the number
density of relevant galaxies is approximately $10^{-2}$~Mpc$^{-3}$.
To get the rate per galaxy, we note that typical initial
mass functions and estimates of the mass needed to evolve into 
a black hole combine to suggest that for a cluster of mass 
$M_{\rm nuc}$, approximately $3\times 10^{-3}(M_{\rm nuc}/M_\odot)$
stars evolve into black holes \citep{OLeary2007}.  
That implies a few$\times 10^4$ black holes
per nuclear star cluster.  If a few tens of percent of these
merge in a Hubble time, and if the rate is slightly lower now 
because many of the original black holes have already merged
(see O'Leary et al. 2006), that suggests a merger rate of
$>0.1\times {\rm few}\times 10^4/(10^{10}~{\rm yr})$
per galaxy, or few$\times 10^{-9}$~Mpc$^{-3}$~yr$^{-1}$.
At the $\sim 2$~Gpc distance at which Advanced LIGO is expected to
be able to see mergers of two $10~M_\odot$ black holes (see,
e.g., \citealt{Mandel07}), 
the available volume is $3\times 10^{10}$~Mpc$^3$, for a rate of
$\gtorder 100$ per year.  Roughly 50--80\% of galaxies in the
eligible luminosity range appear to have nuclear star clusters
(see \citealt{Ferrarese2006} for a summary).  If the majority
of the clusters do not have a supermassive black hole, this 
suggests a final rate of at least several tens per year for Advanced LIGO.
This could be augmented somewhat by small galaxies that
originally had supermassive black holes, but had them ejected after
a merger and then reformed a central cluster 
\citep{Volonteri2007,VHG08}.  

For nearby ($z<0.1$) events of this type it
might be possible to identify the host galaxy.  However, for more typical
$z\sim 0.5\Rightarrow d\approx 2$~Gpc events the number of candidates 
is too large:
even assuming angular localization of $\Delta\Omega=(1^\circ)^2$
and a distance accuracy of $\Delta d/d=1$\%, the number of galaxies 
in the right luminosity range
is $N\sim 4\pi/3 (2000~{\rm Mpc})^3(\Delta\Omega/4\pi)(\Delta d/d)
(0.01~{\rm Mpc}^{-3})\approx 80$.  Therefore, barring some unforseen
electromagnetic counterpart, the host will usually not be obvious.

We anticipate that tens per year is a somewhat
conservative number, because our simulations suggest that more
like 50\% of black holes will be retained, even as single objects,
and because (unlike in a globular cluster) the central regions
of galaxies are not devoid of gas, hence more black holes could
form in the vicinity of the cluster and fall in.  In addition, 
if stellar-mass black holes with masses beyond $20~M_\odot$ are
common, this also increases the detection radius and hence the
rate.  Even for total masses $\sim 30~M_\odot$ and at redshifts
$z\sim 0.5$, the observer
frame gravitational wave frequency at the innermost stable circular
orbit is $f_{\rm ISCO}\sim 4400~{\rm Hz}/[30(1+z)]\sim 100$~Hz.
This is close enough to the range where frequency sensitivity declines
that detection of many of these events will rely strongly on
the signal obtained from the last few orbits plus merger and ringdown.
In much of this range, numerical relativity is essential.

As a final point, we note that for the same reason that nuclear star
clusters are favorable environments for retention and mergers of
stellar-mass black holes, they could also be good birthplaces 
for more massive black holes.  This could be prevented, even for the
relatively high escape speeds discussed here, if recoil from
gravitational radiation during the coalescence exceeds  $\sim
200$~km~s$^{-1}$.  The key uncertainty here is the spin magnitudes of
the holes at birth.  Numerous simulations demonstrate that high spins
with significant projections in the binary orbital plane can produce
kicks of up to several thousand kilometers per second
\citep{Gonzalez07}.  If there is significant
processing of gas through accretion disks the spins are aligned in a
way that reduces the kick to below 200~km~s$^{-1}$
\citep{Bogdanovic2007}, but stellar-mass black holes cannot pick up
enough mass from the interstellar medium for this to be effective. 
For example, the Bondi-Hoyle accretion rate is
${\dot M}_{\rm Bondi}\approx 10^{-13}~M_\odot~{\rm yr}^{-1}
(\sigma/50~{\rm km~s}^{-1})^{-3}(n/100~{\rm cm}^{-3})
(M/10~M_\odot)^2$, meaning that to accrete the $\sim 1$\% of the
black hole mass needed to realign the spin \citep{Bogdanovic2007}
would require at least a trillion years.
Current estimates of stellar-mass black hole spins suggest $a/M>0.5$
in many cases \citep{Shaf06,McClint06,Mil07,Liu2008}.  If the spins are
isotropically oriented and uniformly distributed in the range $0<a/M<1$,
and the mass ratios are in the $m_{\rm small}/m_{\rm big}\sim
0.6-0.8$ range typical in our simulations, then use of the
\cite{Campanelli07} or \cite{Baker08} kick formulae imply that
roughly 84\% of the
recoils exceed 200~km~s$^{-1}$ and 78\% exceed 250~km~s$^{-1}$.  This
suggests that multiple mergers are rare unless there is initially an
extra-massive black hole as a seed (e.g., \citealt{HB08} for a discussion
of the effects of gravitational wave recoil), but further study is important.

In conclusion, we show that the compact nuclear star clusters
found in the centers of many small galaxies are ideal places to foster
mergers between stellar-mass black holes.  It is not clear whether
multiple rounds of mergers can lead to runaway, but this is a new
potential source for ground-based detectors such as Advanced LIGO,
where numerical relativity will play an especially important role.

\acknowledgments

We thank Karl Gebhardt, Kayhan G\"ultekin, Kelly Holley-Bockelmann,
and Jeff McClintock for their many useful suggestions, which helped to
clarify this paper.
This work was supported in part by NASA ATFP grant NNX08AH29G.


\clearpage

\begin{deluxetable}{cccccccc} 
\tablecaption{Simulations of nuclear star clusters\tablenotemark{\rm a}
\label{sims}}
\tablewidth{0pt}
\tablecolumns{8}
\tablehead{
\colhead{$V_{\rm esc}$ (km~s$^{-1}$)\tablenotemark{\rm b}} &
\colhead{$M_{\rm ms,max}$\tablenotemark{\rm c}} &
\colhead{$\alpha$\tablenotemark{\rm d}} &
\colhead{$\langle M_{\rm BH}\rangle (M_\odot)$\tablenotemark{\rm e}} &  
\colhead{$f_{\rm merge}$\tablenotemark{\rm f}} &
\colhead{$f_{\rm not BH}$\tablenotemark{\rm g}} &
\colhead{$\langle M_{\rm bin,merge}\rangle (M_\odot)$\tablenotemark{\rm h}} & 
\colhead{$\langle N_{\rm single,eject}\rangle$\tablenotemark{\rm i}}
}
\startdata
 50 & $1M_\odot$ & 0 & 11.7 & 0.25 & 0.0 & 31.2 & 24.8 \\
62.5& $1M_\odot$ & 0 & 11.7 & 0.33 & 0.0 & 31.6 & 15.3 \\
 75 & $1M_\odot$ & 0 & 11.7 & 0.42 & 0.0 & 30.9 & 11.5 \\
87.5& $1M_\odot$ & 0 & 11.7 & 0.52 & 0.0 & 31.9 & 7.9 \\
 20 & $1M_\odot$ & 0 & 11.7 & 0.63 & 0.02 & 30.0 & 6.2 \\
112.5& $1M_\odot$ & 0 & 11.7 & 0.68 & 0.0 & 31.4 & 4.7 \\
 125 & $1M_\odot$ & 0 & 11.7 & 0.72 & 0.02 & 31.8 & 4.3 \\
137.5& $1M_\odot$ & 0 & 11.7 & 0.76 & 0.01 & 32.0 & 3.0 \\
 150 & $1M_\odot$ & 0 & 11.7 & 0.80 & 0.03 & 32.3 & 2.8 \\
162.5& $1M_\odot$ & 0 & 11.7 & 0.93 & 0.03 & 31.3 & 2.0 \\
 175 & $1M_\odot$ & 0 & 11.7 & 0.89 & 0.02 & 31.9 & 2.0 \\
187.5& $1M_\odot$ & 0 & 11.7 & 0.90 & 0.01 & 31.3 & 2.1 \\
 200 & $1M_\odot$ & 0 & 11.7 & 0.94 & 0.08 & 31.1 & 1.3 \\
212.5& $1M_\odot$ & 0 & 11.7 & 0.89 & 0.05 & 30.5 & 1.0 \\
 225 & $1M_\odot$ & 0 & 11.7 & 0.98 & 0.06 & 31.0 & 1.2 \\
237.5& $1M_\odot$ & 0 & 11.7 & 0.94 & 0.06 & 30.1 & 1.0 \\
 250 & $1M_\odot$ & 0 & 11.7 & 0.96 & 0.06 & 30.0 & 0.71 \\
 200 & $1M_\odot$ & -1.0  & 13.4 & 0.94 & 0 & 32.4   & 1.3 \\
 200 & $1M_\odot$ & -0.5  & 12.6   & 0.95 & 0.01 & 32.2 & 1.5 \\
 200 & $1M_\odot$ & 0.5  & 10.7  & 0.94 & 0.1 & 28.3 & 0.91 \\
 200 & $1M_\odot$ & 1.0  & 9.7 & 0.98 & 0.41 & 27.3 & 0.43 \\
 200 & $1M_\odot$ & 1.5  & 8.8 & 0.99 & 0.79 & 23.0 & 0.04 \\
 200 & $1M_\odot$ & 2.0  & 7.5 & 1.00 & 0.99 & --- & 0 \\
 200 & $1M_\odot$ & 2.35 & 7.4 & 1.00 & 1.00 & ---  & 0 \\
 200 & $3M_\odot$ & -1.0 & 13.4 & 0.85 & 0.03 & 33.3 & 1.5 \\
 200 & $3M_\odot$ & -0.5 & 12.6 & 0.94 & 0.01 & 31.9 & 1.3 \\
 200 & $3M_\odot$ & 0 & 11.7 & 0.95 & 0.05 & 30.4 & 1.5 \\
 200 & $3M_\odot$ & 0.5 & 10.7 & 0.94 & 0.11 & 29.2 & 1.0 \\
 200 & $3M_\odot$ & 1.0 & 9.7 & 0.99 & 0.48 & 25.3 & 0.38 \\
 200 & $3M_\odot$ & 1.5 & 8.8 & 0.99 & 0.85 & 24.7 & 0.04 \\
 200 & $3M_\odot$ & 2.0 & 7.5 & 1.00 & 1.00 & --- & 0 \\
 200 & $3M_\odot$ & 2.35 & 7.4 & 1.00 & 1.00 & --- & 0 \\
\enddata

\tablenotetext{a}{All runs had 100 realizations.}
\tablenotetext{b}{Escape speed from cluster.}
\tablenotetext{c}{Maximum mass of main sequence star.}
\tablenotetext{d}{Number distribution of stars on zero age main
sequence: $dN/dM\propto M^{-\alpha}$.}
\tablenotetext{e}{Average mass of all black holes given $\alpha$ and
our evolutionary assumptions.}
\tablenotetext{f}{Fraction of runs in which holes merged rather than
being ejected.}
\tablenotetext{g}{Fraction of runs in which holes merged with something
other than another black hole.}
\tablenotetext{h}{Average mass of double BH binaries that merged.}
\tablenotetext{i}{Average number of single black holes ejected 
per binary that merged.}
 
\end{deluxetable}


\begin{figure}
\epsscale{1.0} 
\plotone{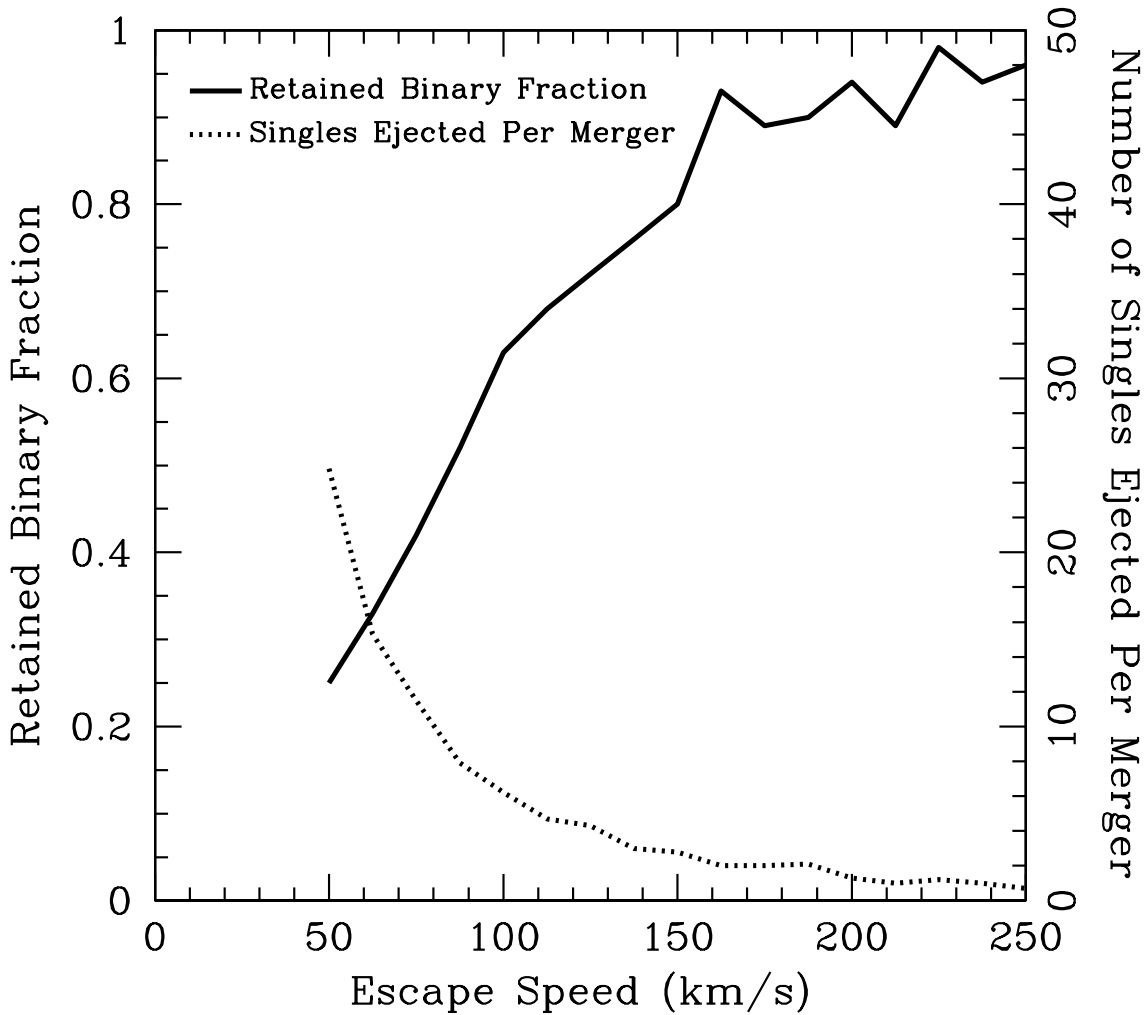}
\figcaption[f1.eps]{Fraction of binaries retained in the nuclear star
cluster (solid line) and average number of black holes ejected per
black hole merger (dotted line) as a function of the cluster escape
speed.  Here the zero age main sequence distribution of masses is
$dN/dM\propto M^0$, to account for mass segregation in the cluster
center, where most interactions occur.  We also assume a maximum black
hole mass of $20~M_\odot$ and a maximum main sequence mass of 
$1~M_\odot$, but most results are robust against variations of these
quantities.  All runs are done with 100 realizations, which explains
the lack of perfect smoothness.  We see, as expected, that the
retention fraction increases rapidly with escape speed, so that for
nuclear star clusters most binaries stay in the cluster until merger.
We also see that at $V_{\rm esc}\sim 200$~km~s$^{-1}$ and above, 
most black hole singles also stay in the cluster.  This suggests
a high merger efficiency.
\label{fig:ejections}}
\end{figure}

\end{document}